\documentclass[journal,10pt,letterpaper]{IEEEtran}
\usepackage[T1]{fontenc}
\usepackage[utf8]{inputenc}
\usepackage[tbtags]{amsmath}
\usepackage{amssymb,bm,graphicx,cite,url}
\usepackage{tikz}
\usetikzlibrary{decorations.pathreplacing}
\usepackage[hidelinks]{hyperref}
\graphicspath{{figures/}}
\begin{document}
\bstctlcite{IEEEcontrol}
\title{Look Before You Move: Multimodal Sensing-Aided Trajectory and Beamforming Design for 6DMA}
\author{Zeyu~Hao, Wentao~Yu,~\IEEEmembership{Member,~IEEE}, Lutz~Lampe,~\IEEEmembership{Senior~Member,~IEEE},\\ and Vincent~W.S.~Wong,~\IEEEmembership{Fellow,~IEEE}%
\thanks{Zeyu Hao, Wentao Yu, Lutz Lampe, and Vincent W.S. Wong are with the Department of Electrical and Computer Engineering, The University of British Columbia, Vancouver, BC, V6T 1Z4, Canada (e-mail: \{zyhao, wentaoyu, lampe, vincentw\}@ece.ubc.ca).}}
\markboth{}{}
\maketitle
\begin{abstract}
Six-dimensional movable antenna (6DMA) technology allows a base station (BS) to adjust the three-dimensional (3D) positions and 3D rotations, i.e., the 6D poses, of its surfaces. Many existing 6DMA designs select the poses based on prior channel knowledge. However, the channels at a new pose can be estimated only after the surfaces have moved to this pose, and the limited movement speeds of the surfaces restrict the poses reachable at the next time step. To address these challenges, we propose a learning-based multimodal sensing-aided trajectory and beamforming design for multiuser 6DMA systems with mobile users and surrounding objects. The BS observes the users and surrounding objects with red-green-blue (RGB) images, light detection and ranging (LiDAR) point clouds, and global positioning system (GPS) positions, independently of the poses of the surfaces. Since a surface may need several time steps to reach a favorable pose, the BS designs a 6DMA trajectory over several future time steps and executes only its first step, following receding-horizon control. We propose a neural network, in which user-centric cross-modal attention and temporal self-attention modules relate each user to the scene and its sensing history, and surface-to-user attention relates each surface to the users. A projection layer keeps the poses feasible under the geometric and kinematic constraints, and a graph attention module allocates the transmit power along the pilot-based beamforming directions. Performance evaluations using the CARLA driving simulator and Sionna ray tracing (RT) demonstrate significant minimum-rate gains of the proposed design over fixed-position antennas. Ablation studies validate the contributions of multimodal sensing and the proposed attention modules to these gains. The results further highlight the importance of designing the trajectory several time steps ahead, which is more beneficial under tighter speed limits of the surfaces.
\end{abstract}
\section{Introduction}
Future wireless networks need to serve an increasing number of users with high spectral efficiency. A common approach is to deploy more antennas at the base station (BS), as in massive multiple-input multiple-output (MIMO) systems, whose hardware cost, energy consumption, and signal processing complexity increase with the array size \cite{massiveBook,10848372}. Existing fixed-position antennas (FPAs) keep their positions and rotations fixed after deployment, allowing the BS to only adapt its transmit signals to the current channel conditions. Movable antennas (MAs) and fluid antennas overcome this limitation by adjusting the positions of individual antennas to reshape the channels \cite{Zhu2024MAModeling,Liao2026FluidMUMIMO,Xu2026MARLMA,10906511,Wong2021FAS,New2025FASTutorial}. Pinching antennas adjust their radiating positions along a waveguide \cite{Xie2026PinchingBlockage,Guo2026GraphTransformerPASS}. The six-dimensional movable antenna (6DMA) architecture further adds rotation as a new degree of freedom, where the BS is equipped with multiple antenna surfaces whose three-dimensional (3D) positions and 3D rotations, i.e., their 6D poses, can be adjusted independently \cite{11142311}. By moving and rotating the surfaces, a 6DMA-BS can turn each surface toward the users and their strongest propagation paths, which can increase the effective channel gains for the users \cite{10752873}. The performance gain of 6DMA depends on how the 6D poses of the surfaces are selected. Selecting these poses involves three main challenges, which we discuss below.

The first challenge in 6DMA is that the channels at a new pose can be estimated only after the surfaces have moved to this pose. The works in \cite{10752873,10848372,stat6dma2025} adapt the poses to the user distribution based on the statistical channel state information (CSI), where the design in \cite{10752873} optimizes continuous 6D poses based on the user distribution subject to geometric constraints on the positions and rotations of the surfaces. Discrete 6D poses are optimized for the average network capacity in \cite{10848372}, and the 6D poses are designed by deep reinforcement learning based on the channels estimated for the candidate poses in \cite{Shao2026HybridNearFar6DMA}. However, the methods described above still rely on statistical CSI or channel estimates at candidate poses to guide pose selection. Thus, selecting favorable poses during online operation without CSI remains a challenge.

The second challenge is that the translation and rotation speeds of the 6DMA surfaces are limited \cite{wangDelay,11142311}. The speed limits, i.e., the kinematic constraints of 6DMA, couple consecutive poses, as the current pose limits the poses reachable at the next time step. A surface may thus need several time steps to turn toward a path that becomes strong later. Since the statistical CSI varies slowly, statistical-CSI-based designs move the surfaces only over long timescales \cite{Shao2025DirSparsity,11142311}. Two-timescale MA designs reposition the antennas only on the slow timescale \cite{KhaliliSchober2026SlowMA,Hu2025TwoTimescaleMAArray}. However, the aforementioned designs \cite{10752873,10848372,stat6dma2025,Shao2025DirSparsity,KhaliliSchober2026SlowMA,Hu2025TwoTimescaleMAArray} ignore this coupling, which can be myopic under changing channels \cite{Cheng2026CKM6DMA}. For MAs, the design in \cite{wangDelay} incorporates the movement delay into throughput maximization based on known angles of the paths. In \cite{Hu2026MATrajectoryFairness}, speed-constrained antenna trajectories are designed to maximize the minimum rate over a transmission period for line-of-sight (LoS) channels with given angles of arrival. Nevertheless, existing approaches either move antennas on a slow timescale or account for the kinematic constraints using given propagation parameters. This calls for 6DMA trajectory design that accounts for how each movement limits the poses reachable at later time steps.

The third challenge is that mobile users and surrounding objects change the propagation environment while the surfaces are moving \cite{Charan2021Blockage,LiYu2026CameraLiDARRF}. As the channels fluctuate rapidly, a pose selected for the average conditions differs from the favorable one at a given time. A pose selected for the current channel may also be outdated by the time the surfaces arrive at this pose \cite{Cheng2026CKM6DMA}. For 6DMA, the design in \cite{Cheng2026CKM6DMA} optimizes 6DMA trajectories over a horizon, but it assumes a perfect channel knowledge map and that user trajectories are known in advance. For vehicular networks, the works in \cite{Ji2026CSIFree6DMAIoV,Ji2026SingleStep6DMA} adjust the 6D poses based on predicted vehicle locations without instantaneous CSI, but they use statistical channel models. The design in \cite{Sun2026MAISACVehicular} tracks the vehicles from sensing echoes to optimize the antenna positions and beamformers, but it considers only LoS channels. The works in \cite{wangDelay,Hu2026MATrajectoryFairness,Cheng2026CKM6DMA,Sun2026MAISACVehicular,Ji2026CSIFree6DMAIoV,Ji2026SingleStep6DMA} rely on prior channel knowledge and do not observe the surrounding objects, which block and create propagation paths as they move. Hence, pose selection requires predicting the future propagation conditions.

Multimodal sensing offers a way to address these three challenges by observing the users and surrounding objects independently of the poses of the surfaces. Since these observations do not require the surfaces to reach the candidate poses, the BS can select future poses before their channels can be estimated. As the sensing history shows how the users and objects move, the BS can also design the poses several time steps ahead under the speed limits. The observations further show how the moving objects change the propagation paths. In the literature, red-green-blue (RGB) images and radar echoes have been used for blockage prediction and vehicle tracking to support proactive beamforming \cite{Charan2021Blockage,Liu2020RadarPredictive}. Multimodal designs fuse RGB images, light detection and ranging (LiDAR) point clouds, and global positioning system (GPS) positions through cross-modal attention \cite{Mollah2026Transformer}, or adapt modality weights through mixture-of-experts models for beam prediction and integrated sensing and communication \cite{Zhang2025MoETHz,Zhang2026MoEISAC}. For multiuser systems, scene observations are fused with pilots for max--min beamforming \cite{11062505,LiYu2026CameraLiDARRF}, or used to estimate propagation directions before digital beamforming \cite{PatelHeath2024}. These sensing-aided designs are all focused on FPAs only and do not design antenna trajectories. Extending them to 6DMA therefore requires selecting future 6D poses under the kinematic constraints while accounting for blockage by moving objects and multiuser interference.

\subsection{Our Contributions}
The research gaps discussed above motivate this question:
\begin{center}
\textit{How can a 6DMA-BS design the trajectory of its surfaces under movement speed limits without CSI?}
\end{center}
To answer this question, we propose a multimodal sensing-aided trajectory and beamforming design for 6DMA systems. The BS first observes the users and surrounding objects and, from their sensing history, designs a 6DMA trajectory over several future time steps before the surfaces move. We refer to this idea as ``look before you move.'' Based on receding-horizon control, it executes only the first step and designs the trajectory again at the next time step. A projection layer maps the candidate poses to poses that satisfy the geometric and kinematic constraints. After the surfaces reach the selected poses, the BS estimates the channels from newly received pilots and uses these estimates to design the beamformers.

To the best of our knowledge, this is the first work to propose a multimodal sensing-aided 6DMA trajectory and beamforming design for multiuser wireless systems. The main contributions of this paper are as follows:
\begin{itemize}
\item We formulate a multimodal sensing-aided 6DMA trajectory and beamforming design for multiuser systems with mobile users and surrounding objects without requiring CSI during the online trajectory design. The formulation maximizes the average minimum rate over the neural network parameters subject to geometric, kinematic, and transmit power constraints. Receding-horizon control accounts for the coupled surface movements by designing the poses several time steps ahead and executing only \mbox{the first step.}
\item We propose an attention-based neural network that relates users to scene features through user-to-RGB attention (R-ATT) and user-to-LiDAR attention (L-ATT), and captures scene evolution through temporal self-attention (T-ATT). Surface-to-user attention (S-ATT) relates each surface to users for trajectory design. A differentiable projection layer maps the candidate poses to poses that satisfy the geometric and kinematic constraints.
\item We propose a graph attention module for power allocation (P-ATT) to design the beamformers at the reached poses. P-ATT combines multimodal features with the estimated user interference on graph edges to allocate transmit power along pilot-based regularized zero-forcing (RZF) directions. We train the trajectory and power-allocation modules jointly to maximize the \mbox{design objective.}
\item Simulations in five traffic scenes built with CARLA and Sionna RT show minimum-rate gains from moving the surfaces over an FPA with the same beamforming. Adding RGB and LiDAR inputs improves upon GPS alone, and ablation studies validate the contribution of every attention module. Our beamforming also achieves a higher minimum rate than two existing multimodal beamforming designs. Longer horizons are more beneficial under tighter speed limits. These results support observing the scene and designing several time steps ahead before the CSI at future poses is available.
\end{itemize}

The rest of this paper is organized as follows. We first discuss the system model in Section~II. Section~III presents the multimodal sensing-aided trajectory and beamforming design for 6DMA systems, i.e., how it operates, what it optimizes, and the neural network that realizes it. Section~IV presents the simulation results. Section~V concludes the paper.

\textit{Notations}: Scalars, vectors, and matrices are denoted by italic letters (e.g., $a$), boldface lowercase letters (e.g., $\bm a$), and boldface uppercase letters (e.g., $\bm A$), respectively. The sets of real and complex numbers are denoted by $\mathbb{R}$ and $\mathbb{C}$, respectively. The transpose, conjugate transpose, and complex conjugate are denoted by $(\cdot)^{\mathsf T}$, $(\cdot)^{\mathsf H}$, and $(\cdot)^*$, respectively. The Euclidean and Frobenius norms are denoted by $\lVert \cdot\rVert_2$ and $\lVert \cdot\rVert_{\mathrm F}$, respectively. The trace, the vertical concatenation of vectors $\bm a$ and $\bm b$, and the \mbox{$k$-th} column of matrix $\bm A$ are denoted by $\operatorname{tr}(\cdot)$, $[\bm a;\bm b]$, and $[\bm A]_{:,k}$, respectively. The $n\times n$ identity matrix, the expectation, and the circularly symmetric complex Gaussian distribution are denoted by $\bm I_n$, $\mathbb{E}[\cdot]$, and $\mathcal{CN}(\cdot,\cdot)$, respectively. Finally, $\mathrm{SO}(3)\triangleq\{\bm{R}\in\mathbb{R}^{3\times3}:\bm{R}^{\mathsf T}\bm{R}=\bm I_3,\det(\bm{R})=1\}$ denotes the special orthogonal (SO) group of degree 3, i.e., the set of 3D rotation matrices.
\section{System Model}
\begin{figure}[!t]
\centering\includegraphics[width=\columnwidth]{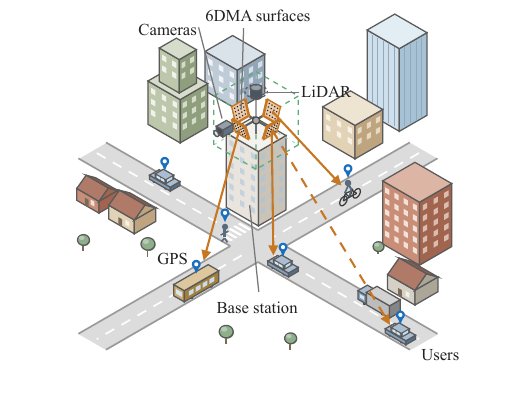}
\caption{The multiuser downlink system with 6DMA surfaces, RGB and LiDAR sensing at the BS, and GPS positions reported by the users. The dashed cube indicates the site space, i.e., the region in which the surface centers can move. Solid and dashed arrows denote unblocked and blocked signals, respectively.}
\label{fig:scene}
\end{figure}

We consider the multiuser downlink 6DMA system as shown in Fig.~\ref{fig:scene}, in which a BS with $B$ 6DMA surfaces serves $K$ users. Let $\mathcal B\triangleq\{1,\ldots,B\}$ and $\mathcal K\triangleq\{1,\ldots,K\}$ denote the sets of 6DMA surfaces and users, respectively. The BS senses the environment with multiple RGB cameras and a LiDAR sensor, and each user reports its GPS position to the BS. Each surface has $N_{\mathrm{sur}}$ antenna elements, and the 6DMA-BS has $N=BN_{\mathrm{sur}}$ antennas in total. All surfaces are connected to a central processing unit (CPU) via extendable and rotatable rods, and the CPU controls the motors on each rod to move and rotate each surface independently. Since the users and objects move, we divide the operation into time steps indexed by $t$, where two consecutive time steps are $\Delta_{\mathrm s}$ apart. At time step $t$, the BS estimates the channels from the pilots of the users, serves the users, and obtains a new sensing frame of RGB images, a LiDAR point cloud, and GPS positions. We model the system at these time steps.

\subsection{6DMA Movement Model}\label{sec:system}
\begin{figure}[!t]
\centering\includegraphics[width=\columnwidth]{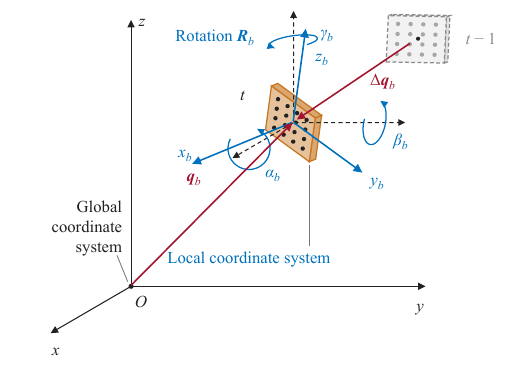}
\caption{Global and local coordinate systems, where the rotations by $\alpha_b$, $\beta_b$, and $\gamma_b$ about the three axes through the surface center illustrate $\bm{R}_b$, and $\Delta\bm{q}_b$ is the translation of surface $b$ between two consecutive time steps.}
\label{fig:geometry}
\end{figure}

We adopt the 6DMA movement model and the geometric constraints of \cite{10752873}. Since the BS changes the poses at every time step, we add the kinematic constraints of \cite{Cheng2026CKM6DMA}. For surface $b\in\mathcal B$, let $\bm{q}_b\in\mathbb{R}^3$ and $\bm{R}_b\in\mathrm{SO}(3)$ denote its center position and rotation, respectively, in a global coordinate system whose origin $O$ is at the CPU. The position $\bm{q}_b$ and the rotation $\bm{R}_b$ form the 6D pose $(\bm{q}_b,\bm{R}_b)$ of surface $b$. With the rotation angles about the \mbox{$x$-}, \mbox{$y$-}, and \mbox{$z$-axes} being the Euler angles $\alpha_b$, $\beta_b$, and $\gamma_b$, respectively, the rotation matrix is given by $\bm R_b=\bm R(\alpha_b,\beta_b,\gamma_b)$, with
\begin{equation}
\setlength{\arraycolsep}{2pt}
\begin{split}
\bm R(\alpha,\beta,\gamma)\triangleq{}&
\begin{bmatrix}1&0&0\\0&\cos\alpha&\sin\alpha\\0&-\sin\alpha&\cos\alpha\end{bmatrix}
\begin{bmatrix}\cos\beta&0&-\sin\beta\\0&1&0\\\sin\beta&0&\cos\beta\end{bmatrix}\\
&\times\begin{bmatrix}\cos\gamma&\sin\gamma&0\\-\sin\gamma&\cos\gamma&0\\0&0&1\end{bmatrix}.
\end{split}
\label{eq:rotation_matrix}
\end{equation}
As illustrated in Fig.~\ref{fig:geometry}, each surface lies in the $y_b$--$z_b$ plane of its local coordinate system, and its outward normal points along the $x_b$ axis of this system. Let $\bar{\bm r}_{b,n}\in\mathbb{R}^3$ denote the position of antenna element $n\in\{1,\ldots,N_{\mathrm{sur}}\}$ on surface $b$ in this local coordinate system. Then, the global position of antenna element $n$ and the outward normal of surface $b$ are respectively given by
\begin{align}
\bm r_{b,n}&=\bm{q}_b+\bm{R}_b\bar{\bm r}_{b,n},\label{eq:element}\\
\bm n_b&=\bm{R}_b[1,0,0]^{\mathsf T}.\label{eq:normal}
\end{align}
We denote the poses of all surfaces by $\bm x\triangleq\{(\bm{q}_b,\bm{R}_b)\}_{b\in\mathcal B}$, and the poses at time step $t$ by $\bm x_t$.

Let $\mathcal C\subset\mathbb{R}^3$ denote the site space, a bounded convex region in which the surface centers can be placed. Let $d_{\min}$ denote the minimum distance between two surface centers. The 6D poses should satisfy the following geometric constraints of \cite{10752873}. Each surface center lies in $\mathcal C$, any two surface centers are at least $d_{\min}$ apart, no two surfaces face each other, and each surface faces outward. Practical motors move each surface at limited speeds. Let $v_{\max}$ and $\omega_{\max}$ denote the translation and rotation speed limits, respectively. Any two consecutive poses should satisfy the kinematic constraints, i.e., within one time step, the center of a surface moves by at most $v_{\max}\Delta_{\mathrm s}$, and the surface rotates by at most the angle $\omega_{\max}\Delta_{\mathrm s}$, where the rotation angle from $\bm{R}$ to $\bm{R}'$ is given by
\begin{equation}
d_{\mathrm R}(\bm{R},\bm{R}')\triangleq
\arccos\!\left(\frac{\operatorname{tr}(\bm{R}^{\mathsf T}\bm{R}')-1}{2}\right).
\label{eq:rotation_distance}
\end{equation}
The kinematic constraints couple consecutive poses, since the current poses limit the poses that can be reached at the next time step. The optimization problem in Section~\ref{sec:problem} includes all constraints.

\subsection{Channel and Signal Models}
As in \cite{10752873}, we model the channel by the propagation paths of each user and the directional radiation pattern of the antenna elements, with multiple paths per user, which change as the moving objects block and create paths. Let $\bm{h}_{k,t}(\bm x)\in\mathbb{C}^N$ denote the narrowband downlink channel from the BS to user $k$ at time step $t$ when the surfaces are at poses $\bm x$. We model the entry of $\bm{h}_{k,t}(\bm x)$ for antenna element $n$ of surface $b$, i.e., entry $(b-1)N_{\mathrm{sur}}+n$ of $\bm{h}_{k,t}(\bm x)$, as
\begin{equation}
\begin{split}
[\bm{h}_{k,t}(\bm x)]_{b,n}
=\sum_{\ell\in\mathcal L_{k,t}} a_{k,\ell,t}\,
g_{b,k,\ell,t}(\bm{R}_b)\,\\
\times e^{-\mathrm j2\pi D_{b,n,k,\ell,t}(\bm{q}_b,\bm{R}_b)/\lambda},
\end{split}
\label{eq:channel}
\end{equation}
where $\mathcal L_{k,t}$ is the set of propagation paths from the BS to user $k$ at time step $t$, $a_{k,\ell,t}$ is the complex coefficient of path $\ell\in\mathcal L_{k,t}$ at time step $t$, including the path loss and reflection coefficients, $g_{b,k,\ell,t}(\bm{R}_b)$ is the effective antenna gain of surface $b$ for this path in amplitude, $D_{b,n,k,\ell,t}(\bm{q}_b,\bm{R}_b)$ is the propagation distance of this path from antenna element $n$ of surface $b$ at $\bm r_{b,n}$ in \eqref{eq:element} to user $k$, and $\lambda$ is the carrier wavelength. Each antenna element has a directional radiation pattern with $g_{b,k,\ell,t}(\bm{R}_b)=10^{A(\vartheta_{b,k,\ell,t}(\bm{R}_b),\varphi_{b,k,\ell,t}(\bm{R}_b))/20}$, where $\vartheta_{b,k,\ell,t}(\bm{R}_b)$ and $\varphi_{b,k,\ell,t}(\bm{R}_b)$ are the elevation and azimuth angles of departure of path $\ell$ to user $k$ at time step $t$, respectively, measured from the outward normal of surface $b$, and the effective antenna gain in dBi is given by \cite{tr38901}
\begin{equation}
A(\vartheta,\varphi)=G_{\max}-\min\Big\{12\Big(\frac{\vartheta}{\vartheta_{\mathrm{3dB}}}\Big)^2+12\Big(\frac{\varphi}{\varphi_{\mathrm{3dB}}}\Big)^2,A_{\mathrm m}\Big\},
\label{eq:pattern}
\end{equation}
where $\vartheta_{\mathrm{3dB}}$ and $\varphi_{\mathrm{3dB}}$ are the \mbox{3-dB} beamwidths in elevation and azimuth, respectively, $G_{\max}$ is the maximum directional gain in dBi, and $A_{\mathrm m}$ is the front-to-back ratio in dB. By \eqref{eq:pattern}, the effective antenna gain of a surface for a path decreases as the path deviates farther from the outward normal. Hence, a surface provides a strong channel for user $k$ only when it faces the paths for user $k$ with large path coefficients. The channel depends on the poses through the effective antenna gains and the propagation distances, and on the environment through $\mathcal L_{k,t}$, which changes as the users and objects move.

At time step $t$, let $s_{k,t}\sim\mathcal{CN}(0,1)$ denote the data symbol for user $k$, where the symbols of different users are independent, and let $\bm{w}_{k,t}\in\mathbb{C}^N$ denote the transmit beamformer for user $k$. We define $\bm{W}_t\triangleq[\bm{w}_{1,t},\ldots,\bm{w}_{K,t}]\in\mathbb{C}^{N\times K}$. The total transmit power of the beamformers is limited by the transmit power budget $P_{\mathrm T}$. The received signal at user $k$ at time step $t$ is given by
\begin{equation}
y_{k,t}=\underbrace{\bm{h}_{k,t}^{\mathsf H}(\bm x_t)\bm{w}_{k,t}s_{k,t}}_{\text{desired signal}}
+\underbrace{\sum_{i\in\mathcal K\setminus\{k\}}\bm{h}_{k,t}^{\mathsf H}(\bm x_t)\bm{w}_{i,t}s_{i,t}}_{\text{multiuser interference}}
+\underbrace{z_{k,t}}_{\text{noise}},
\label{eq:received}
\end{equation}
where $z_{k,t}\sim\mathcal{CN}(0,\sigma^2)$ is the additive white Gaussian noise at user $k$. The signal-to-interference-plus-noise ratio (SINR) and the achievable rate of user $k$ are respectively given by
\begin{align}
\Gamma_{k,t}(\bm x_t,\bm{W}_t)&=
\frac{|\bm{h}_{k,t}^{\mathsf H}(\bm x_t)\bm{w}_{k,t}|^2}
{\sigma^2+\sum_{i\in\mathcal K\setminus\{k\}}|\bm{h}_{k,t}^{\mathsf H}(\bm x_t)\bm{w}_{i,t}|^2},\label{eq:sinr}\\
R_{k,t}(\bm x_t,\bm{W}_t)&=\log_2(1+\Gamma_{k,t}(\bm x_t,\bm{W}_t)).\label{eq:rate}
\end{align}

\subsection{Multimodal Sensing and Channel Estimation}\label{sec:estimation}
In addition to the pilots, the BS uses the RGB images, the LiDAR point clouds, and the GPS positions. As the channels at a new pose can be estimated only after the surfaces have moved to it, the BS relies on these sensing data to select the next poses. The RGB images show which objects are present, the LiDAR point clouds capture their 3D geometry, and the GPS positions locate the users. At time step $t$, the multimodal sensing data available at the BS are given by
\begin{equation}
\begin{split}
\mathcal I_t\triangleq
\{\bm C_{t'},\bm L_{t'},\widehat{\bm p}_{k,t'}\}_{k\in\mathcal K,\,t'\in\mathcal H_t},
\end{split}
\label{eq:information}
\end{equation}
where $\mathcal H_t\triangleq\{t-H+1,\ldots,t\}$ is the set of the latest $H$ time steps, $\bm C_{t'}$ and $\bm L_{t'}$ are the RGB images and the LiDAR point cloud captured at time step $t'$, respectively, and $\widehat{\bm p}_{k,t'}\in\mathbb{R}^3$ is the GPS position reported by user $k$ at time step $t'$ in the global coordinate system.

While the multimodal sensing data $\mathcal I_t$ describe the scene, the SINRs in \eqref{eq:sinr} depend on the channels at the reached poses. Since the channels change whenever the surfaces move, the users transmit fresh pilots at each time step $t$ after the surfaces reach the poses $\bm x_t$, and the BS estimates the channels at $\bm x_t$ from these pilots. We consider the time-division duplex (TDD) mode, in which the downlink channels are obtained from the uplink pilots by channel reciprocity. All users transmit their uplink pilots at the same time over $\tau_{\mathrm p}$ symbols. Let $\bm\phi_k\in\mathbb{C}^{\tau_{\mathrm p}}$ with $\lVert \bm\phi_k\rVert_2^2=\tau_{\mathrm p}$ denote the pilot sequence from user $k$. We assume that the pilot sequences of different users are mutually orthogonal, i.e., $\bm\phi_i^{\mathsf T}\bm\phi_k^{*}=0$ for $i\neq k$, which requires $\tau_{\mathrm p}\geq K$. The pilot signal received at the BS at time step $t$ is given by
\begin{equation}
\bm Y_t=\sqrt{P_{\mathrm p}}\sum_{k\in\mathcal K}\bm{h}_{k,t}(\bm x_t)\bm\phi_k^{\mathsf T}+\bm E_{\mathrm p,t},
\label{eq:pilot}
\end{equation}
where $P_{\mathrm p}$ is the pilot power of each user, and the entries of the pilot noise $\bm E_{\mathrm p,t}\in\mathbb{C}^{N\times\tau_{\mathrm p}}$ are independent $\mathcal{CN}(0,\sigma_{\mathrm p}^2)$. The BS correlates $\bm Y_t$ with $\bm\phi_k$ and obtains the least-squares (LS) channel estimate of user $k$ at time step $t$ as
\begin{equation}
\widehat{\bm h}_{k,t}
\triangleq\frac{\bm Y_t\bm\phi_k^{*}}{\tau_{\mathrm p}\sqrt{P_{\mathrm p}}},
\label{eq:ls}
\end{equation}
where $\widehat{\bm h}_{k,t}-\bm h_{k,t}(\bm x_t)\sim\mathcal{CN}(\bm 0,\sigma_{\mathrm p}^2(\tau_{\mathrm p}P_{\mathrm p})^{-1}\bm I_N)$ is the channel estimation error.
\section{Proposed Multimodal Sensing-Aided Trajectory and Beamforming Design}
In this section, we present an overview and the objective of the proposed design, and then the network that realizes it.

\subsection{Overview of the Proposed Design}\label{sec:concept}
\begin{figure}[!t]
\centering
\begin{tikzpicture}[x=0.86cm,y=0.66cm,font=\footnotesize]
\draw[->,thick] (-0.4,0)--(9.2,0);
\node[above left] at (9.2,0.03) {time step};
\foreach \x/\l in {0/{t-H+1},3/{t},4.4/{t+1},5.9/{t+f},7.4/{t+F}}{\draw[thick] (\x,0.09)--(\x,-0.09) node[below]{$\l$};}
\node at (1.5,-0.27) {$\cdots$};
\draw[decorate,decoration={brace,amplitude=4pt}] (0,0.22)--(3,0.22) node[midway,above=4pt,align=center]{latest $H$\\sensing frames};
\draw[decorate,decoration={brace,amplitude=4pt}] (4.4,0.22)--(7.4,0.22) node[midway,above=4pt,align=center]{6D poses designed\\at time step $t$};
\fill (3,0) circle (1.6pt);
\node[below left,inner sep=1pt] at (3.25,-0.62) {current poses $\bm x_t$};
\node[below,inner sep=1pt] at (4.4,-0.62) {executed};
\end{tikzpicture}
\caption{Time steps involved in the design at time step $t$.}
\label{fig:timeline}
\end{figure}
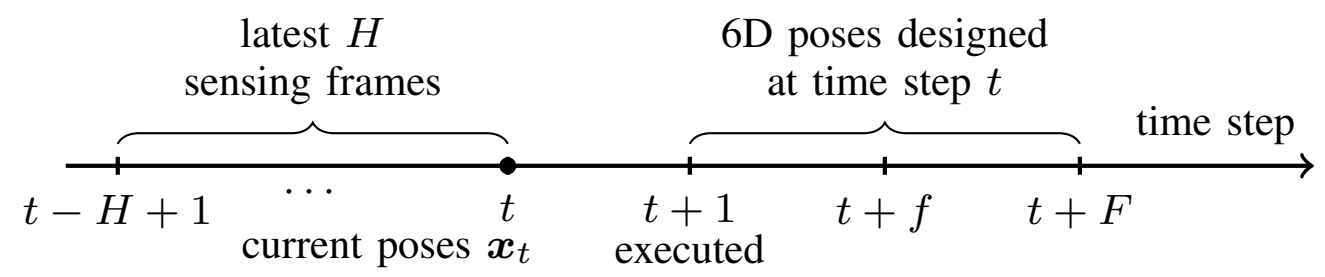

Since the kinematic constraints couple consecutive poses, we adopt the receding-horizon principle of model predictive control, in which the BS designs the poses for several time steps ahead but executes only the first designed step. At time step $t$, the BS designs the 6D poses for the next $F$ time steps $t+1,\ldots,t+F$ based on the latest $H$ sensing frames, where we refer to $F$ as the horizon. Let $\mathcal F\triangleq\{1,\ldots,F\}$ denote the set of steps in the horizon, and let $\bm x_{t,f}\triangleq\{(\bm{q}_{b,t,f},\bm{R}_{b,t,f})\}_{b\in\mathcal B}$ denote the 6D poses designed at time step $t$ for time step $t+f$, $f\in\mathcal F$, where we define $\bm x_{t,0}\triangleq\bm x_t$, so that $(\bm{q}_{b,t,0},\bm{R}_{b,t,0})$ is the current pose of surface $b$. We refer to $\bm x_{t,1:F}\triangleq(\bm x_{t,1},\ldots,\bm x_{t,F})$ as the 6DMA trajectory designed at time step $t$. At time step $t+1$, the BS designs the trajectory again from the reached poses $\bm x_{t+1}=\bm x_{t,1}$ with the new sensing frame. Fig.~\ref{fig:timeline} illustrates these time steps.

We implement this design by a neural network that processes the multimodal sensing data, the current poses, and the received pilots. Let $\{\bm\theta,\bm\psi\}$ denote the trainable parameters of the entire neural network, which consists of the trajectory design module with parameters $\bm\theta$, including the multimodal history encoding in Section~\ref{sec:encoding}, and the power-allocation network with parameters $\bm\psi$. The power-allocation network also uses the features of this encoding. At each time step $t$, the trajectory design module $\pi_{\bm\theta}$ and the beamforming mapping $\mathcal W_{\bm\theta,\bm\psi}$ operate as
\begin{align}
\bm x_{t,1:F}&=\pi_{\bm\theta}(\mathcal I_t,\bm x_t),\label{eq:plan}\\
\bm x_{t+1}&=\bm x_{t,1},\label{eq:execute}\\
\bm{W}_t&=\mathcal W_{\bm\theta,\bm\psi}(\mathcal I_{t-1},\bm x_t,\bm Y_t).\label{eq:beammap}
\end{align}
The beamforming mapping uses $\mathcal I_{t-1}$, since the sensing frame of time step $t$ is obtained after the users are served.

\begin{figure*}[!t]
\centering\includegraphics[width=\textwidth]{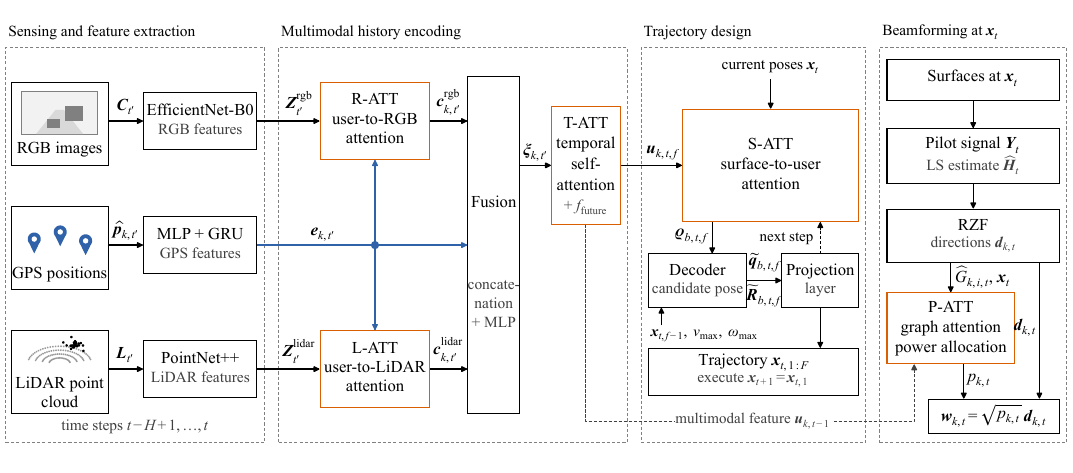}
\caption{Architecture of the proposed multimodal sensing-aided trajectory and beamforming design for 6DMA.}
\label{fig:architecture}
\end{figure*}
\subsection{Design Objective}\label{sec:problem}
We use the minimum rate, rather than the sum rate, as the design objective. The sum rate can be increased by allocating more power to the users with strong channels, whereas the minimum rate depends on the users with weak channels, e.g., the users whose strongest paths are blocked by moving objects. Maximizing the minimum rate thus requires the BS to keep these users served as the objects move, which motivates observing the objects by multimodal sensing. Since the design is a mapping that the BS applies at every time step, we optimize the parameters of this mapping rather than the poses of one scene. Under the geometric constraints, the kinematic constraints, and the transmit power constraint, we maximize the average minimum rate over $T$ time steps as
\begin{subequations}\label{eq:problem}
\begin{align}
\underset{\bm\theta,\bm\psi}{\text{maximize}}\ \ &\mathbb{E}\!\left[\frac1T\sum_{t\in\mathcal T}\min_{k\in\mathcal K}R_{k,t}(\bm x_t,\bm{W}_t)\right]\\
\text{subject to}\ \ &\bm{q}_{b,t,f}\in\mathcal C,\ \ b\in\mathcal B,\ f\in\mathcal F,\ t\in\mathcal T,\label{eq:box}\\
&\lVert \bm{q}_{b,t,f}-\bm{q}_{b',t,f}\rVert_2\geq d_{\min},\notag\\
&\qquad b\in\mathcal B,\ b'\in\mathcal B\setminus\{b\},\ f\in\mathcal F,\ t\in\mathcal T,\label{eq:separation}\\
&\bm n_{b,t,f}^{\mathsf T}(\bm{q}_{b',t,f}-\bm{q}_{b,t,f})\leq0,\notag\\
&\qquad b\in\mathcal B,\ b'\in\mathcal B\setminus\{b\},\ f\in\mathcal F,\ t\in\mathcal T,\label{eq:nonfacing}\\
&\bm n_{b,t,f}^{\mathsf T}\bm{q}_{b,t,f}\geq0,\ \ b\in\mathcal B,\ f\in\mathcal F,\ t\in\mathcal T,\label{eq:outward}\\
&\lVert \bm{q}_{b,t,f}-\bm{q}_{b,t,f-1}\rVert_2\leq v_{\max}\Delta_{\mathrm s},\notag\\
&\qquad b\in\mathcal B,\ f\in\mathcal F,\ t\in\mathcal T,\label{eq:step_position}\\
&d_{\mathrm R}(\bm{R}_{b,t,f-1},\bm{R}_{b,t,f})\leq\omega_{\max}\Delta_{\mathrm s},\notag\\
&\qquad b\in\mathcal B,\ f\in\mathcal F,\ t\in\mathcal T,\label{eq:step_rotation}\\
&\lVert \bm{W}_t\rVert_{\mathrm F}^2\leq P_{\mathrm T},\ \ t\in\mathcal T,\label{eq:power}
\end{align}
\end{subequations}
where $\mathcal T\triangleq\{1,\ldots,T\}$ is the set of time steps, $\bm n_{b,t,f}\triangleq\bm{R}_{b,t,f}[1,0,0]^{\mathsf T}$ is the outward normal of surface $b$ at the designed pose $(\bm{q}_{b,t,f},\bm{R}_{b,t,f})$ by \eqref{eq:normal}, $\bm x_t$ and $\bm W_t$ follow \eqref{eq:plan}--\eqref{eq:beammap}, and the expectation is taken over the random motion of the users and objects and the pilot noise $\bm E_{\mathrm p,t}$. Constraint \eqref{eq:box} keeps each surface center inside the site space, and constraint \eqref{eq:separation} avoids the overlap and mutual coupling of the surfaces. Constraint \eqref{eq:nonfacing} prevents any two surfaces from facing each other to avoid mutual signal reflections. Constraint \eqref{eq:outward} makes each surface face outward to avoid signal blockage. Constraints \eqref{eq:step_position} and \eqref{eq:step_rotation} are the kinematic constraints between consecutive time steps, and constraint \eqref{eq:power} is the transmit power constraint.

\subsection{Motivation of the Learning-Based Design}\label{sec:motivation}
Since the channels at the designed poses can be estimated only after the surfaces have moved to these poses, we decompose the design in problem~\eqref{eq:problem} into two subproblems. The trajectory design forms the outer subproblem, whose objective evaluates the designed poses of all $F$ steps by the minimum rate achieved by the optimal beamformers at these poses. Let $R_{k,t+f}^\star(\bm x_{t,f})$ denote the achievable rate of user $k$ at time step $t+f$ at the designed poses $\bm x_{t,f}$ with the beamformers that maximize the minimum rate for the true channels at these poses. The trajectory design subproblem is given by
\begin{subequations}\label{eq:pose_sub}
\begin{align}\underset{\bm x_{t,1:F}}{\text{maximize}}\quad&\mathbb{E}\!\left[\sum_{f\in\mathcal F}\min_{k\in\mathcal K}R_{k,t+f}^\star(\bm x_{t,f})\,\Big|\,\mathcal I_t\right]\\\text{subject to}\quad&\text{constraints \eqref{eq:box}--\eqref{eq:step_rotation} at time step $t$},\end{align}
\end{subequations}
where $R_{k,t+f}^\star(\bm x_{t,f})$ depends on the channels at the future time step $t+f$. Problem~\eqref{eq:pose_sub} is difficult to solve for two reasons. First, the channels at the future time steps are unknown when the BS designs the poses. Second, problem~\eqref{eq:pose_sub} is non-convex, since the channel in \eqref{eq:channel} depends on the poses through the phases of the paths and through the effective antenna gains in \eqref{eq:pattern}, and the set $\mathrm{SO}(3)$ of rotations is non-convex. Therefore, we learn the trajectory design module $\pi_{\bm\theta}$ that maps $\mathcal I_t$ and $\bm x_t$ to the designed poses.

For the reached poses $\bm x_t$, by introducing an auxiliary variable $R_{\min}$ for the minimum rate, we can express the inner beamforming subproblem at time step $t$ as
\begin{subequations}\label{eq:bf_sub}
\begin{align}\underset{R_{\min},\,\bm{W}_t}{\text{maximize}}\quad&R_{\min}\\\text{subject to}\quad&R_{k,t}(\bm x_t,\bm{W}_t)\geq R_{\min},\quad k\in\mathcal K,\\&\lVert \bm{W}_t\rVert_{\mathrm F}^2\leq P_{\mathrm T}.\end{align}
\end{subequations}
Given the channels $\bm{h}_{k,t}(\bm x_t)$, $k\in\mathcal K$, problem~\eqref{eq:bf_sub} can be solved by a bisection search over an SINR target, where each iteration solves a second-order cone program \cite{wiesel_conic}. However, this algorithm has two limitations in online operation. First, the algorithm can use only the channel estimates in \eqref{eq:ls}, whose errors degrade its solution for the true channels. In contrast, the multimodal sensing data describe the large-scale fading effects of each user, such as path loss and blockage, and can be fused with the pilots to improve the beamformers. Second, the algorithm needs to solve a sequence of convex problems at each time step, whereas a trained neural network computes its output by only one forward pass. Hence, we learn the beamforming mapping $\mathcal W_{\bm\theta,\bm\psi}$ that maps the channel estimates, the multimodal features, and the current poses to the beamformers. We train the whole network in an end-to-end unsupervised manner, i.e., using the minimum-rate objective without labels, since an optimal 6DMA trajectory label would require solving the non-convex problem~\eqref{eq:pose_sub} for \mbox{every training sample.}

The design of our proposed network follows three intrinsic properties of the sensing data and problem~\eqref{eq:problem}. First, the BS observes the scene in each frame as a set of features, i.e., image regions in the RGB images and local regions of points in the LiDAR point cloud. The order of these features is irrelevant. The importance of a feature to a user depends on its position relative to the user. Second, the users have no natural order. If the users are indexed in a different order, the designed poses remain unchanged, and the beamformers are permuted in the same way as the users. Third, the users interfere with each other, as shown in \eqref{eq:sinr}, so that the power allocated to one user affects the SINRs of all other users. Attention is suitable for the first two properties, since it computes, for each query, a weighted sum of the inputs that is invariant to permutations of the inputs.

As shown in Fig.~\ref{fig:architecture}, the network processes the data in three stages, i.e., the multimodal history encoding in Section~\ref{sec:encoding}, the trajectory design in Sections~\ref{sec:trajectory} and \ref{sec:recovery}, and the pilot-based beamforming in Section~\ref{sec:beamforming}. R-ATT and L-ATT exploit the first property, S-ATT and the parameters shared by all users exploit the second property, and the interference graph in P-ATT exploits the third property.

\subsection{Multimodal History Encoding}\label{sec:encoding}
The goal of the encoding is to obtain, for each user, a feature that summarizes the user and its surroundings over the latest $H$ historical time steps, since the future channel of the user depends on both its motion and the objects around it.

\subsubsection{Modality-Specific Encoders}
GPS positions describe how the users move, whereas RGB images and LiDAR point clouds describe the objects that determine the propagation environment. For the RGB images and the LiDAR point clouds, we use the EfficientNet-B0 encoder in \cite{tan2019efficientnet} and the PointNet++ encoder in \cite{pointnetpp}, respectively. For each time step $t'\in\mathcal H_t$, they extract the RGB features $\bm Z_{t'}^{\mathrm{rgb}}\in\mathbb{R}^{M_{\mathrm r}\times d}$ and the LiDAR features $\bm Z_{t'}^{\mathrm{lidar}}\in\mathbb{R}^{M_{\mathrm l}\times d}$ of dimension $d$ from the RGB images and the point cloud, respectively, where $M_{\mathrm r}$ and $M_{\mathrm l}$ are the numbers of RGB features and LiDAR features, respectively. We denote the $j$-th feature of modality $m\in\{\mathrm{rgb},\mathrm{lidar}\}$ at time step $t'$ by $\bm z_{j,t'}^{m}$. The RGB encoder outputs a feature map for the image of each camera, and each RGB feature describes one region of one image. Since the camera poses are known, each region corresponds to a known direction from its camera. Each LiDAR feature describes a local region of points, whose centroid is a known position in space. We map the GPS positions of user $k$ up to time step $t'$ to the GPS feature $\bm e_{k,t'}\in\mathbb{R}^d$ by a GPS encoder, which is shared by all users and consists of a multilayer perceptron (MLP) followed by a gated recurrent unit (GRU). The GRU extracts the temporal feature of the user movement.

\subsubsection{User-Centric Cross-Modal Attention}
Since the channel of each user depends on the objects in its propagation environment, R-ATT and L-ATT collect, for each user, the RGB and LiDAR features that describe these objects, respectively. In both modules, the GPS feature of user $k$ forms the query, and the features of the modality form the keys and values. The output of R-ATT or L-ATT for user $k$ and modality $m$ at time step $t'$ is given by
\begin{equation}
\bm c_{k,t'}^{m}=\sum_{j\in\mathcal V_{t'}^m}\zeta_{k,j,t'}^{m}\bm\Omega_{\mathrm V}^m\bm z_{j,t'}^{m},
\label{eq:spatial_context}
\end{equation}
with the attention weight being
\begin{align}
\zeta_{k,j,t'}^{m}&=\frac{\exp(\epsilon_{k,j,t'}^{m})}{\sum_{j'\in\mathcal V_{t'}^m}\exp(\epsilon_{k,j',t'}^{m})},\label{eq:spatial_attention_w}\\
\epsilon_{k,j,t'}^{m}&=\frac{(\bm\Omega_{\mathrm Q}^m\bm e_{k,t'})^{\mathsf T}\bm\Omega_{\mathrm K}^m\bm z_{j,t'}^{m}}{\sqrt{d_{\mathrm h}}}+\delta_{k,j,t'}^{m},\label{eq:spatial_attention}
\end{align}
where $\epsilon_{k,j,t'}^{m}$ is the attention score, $\zeta_{k,j,t'}^{m}$ measures the importance of feature $j$ to user $k$, $\bm\Omega_{\mathrm Q}^m$, $\bm\Omega_{\mathrm K}^m$, and $\bm\Omega_{\mathrm V}^m$ are the trainable query, key, and value projection matrices, respectively, $\mathcal V_{t'}^m$ is the index set of the features of modality $m$ at time step $t'$, and $d_{\mathrm h}$ is the dimension of each projected query and key. To help the attention locate user $k$, we include the geometric bias $\delta_{k,j,t'}^{m}$ in the attention score $\epsilon_{k,j,t'}^{m}$ in \eqref{eq:spatial_attention} \cite{gao2021smca}. The bias depends on the offset between the known location of feature $j$ and the reported position $\widehat{\bm p}_{k,t'}$ of user $k$. For R-ATT, the offset is the angle between the direction of image region $j$ and the direction from its camera to $\widehat{\bm p}_{k,t'}$. For L-ATT, the offset is the distance between the centroid of the local region of feature $j$ and $\widehat{\bm p}_{k,t'}$. The bias $\delta_{k,j,t'}^{m}$ decreases quadratically with this offset at a trainable positive rate.
\subsubsection{Multimodal Fusion}
The three modalities describe user $k$ from complementary views, i.e., its position from GPS, and the classes and the 3D geometry of the objects around it from RGB and LiDAR, respectively. We concatenate the outputs of R-ATT and L-ATT with the GPS feature of user $k$ at time step $t'$ and fuse them as $\bm\xi_{k,t'}=f_{\mathrm{fuse}}([\bm e_{k,t'};\bm c_{k,t'}^{\mathrm{rgb}};\bm c_{k,t'}^{\mathrm{lidar}}])$, where $f_{\mathrm{fuse}}$ is an MLP with output dimension $d$.

\subsubsection{Temporal Self-Attention and Future User Features}
A single frame shows where the user and the objects around it are, but not how the objects move. Hence, T-ATT relates the fused features of each user across the $H$ frames, with the same parameters for all users. For user $k$, T-ATT applies self-attention to the fused features $[\bm\xi_{k,t-H+1},\ldots,\bm\xi_{k,t}]$ of the time steps in $\mathcal H_t$, where the fused feature at time step $t$ forms the query. Since attention alone ignores the order of the frames, we add the sinusoidal positional encoding of time step $t'$ to each fused feature $\bm\xi_{k,t'}$ and obtain $\bar{\bm\xi}_{k,t'}$. The output of T-ATT for user $k$ at the latest time step $t$ is given by
\begin{equation}
\bm u_{k,t}=\sum_{t'\in\mathcal H_t}\zeta_{k,t,t'}^{\mathrm{time}}\bm\Omega_{\mathrm V}^{\mathrm{time}}\bar{\bm\xi}_{k,t'},
\label{eq:tatt}
\end{equation}
with the attention weight being
\begin{align}
\zeta_{k,t,t'}^{\mathrm{time}}&=\frac{\exp(\epsilon_{k,t,t'}^{\mathrm{time}})}{\sum_{t''\in\mathcal H_t}\exp(\epsilon_{k,t,t''}^{\mathrm{time}})},\label{eq:tatt_weight_w}\\
\epsilon_{k,t,t'}^{\mathrm{time}}&=\frac{(\bm\Omega_{\mathrm Q}^{\mathrm{time}}\bar{\bm\xi}_{k,t})^{\mathsf T}\bm\Omega_{\mathrm K}^{\mathrm{time}}\bar{\bm\xi}_{k,t'}}{\sqrt{d_{\mathrm h}}},\label{eq:tatt_weight}
\end{align}
where $\epsilon_{k,t,t'}^{\mathrm{time}}$ is the attention score, $\zeta_{k,t,t'}^{\mathrm{time}}$ measures the importance of the feature of user $k$ at time step $t'$ to that at time step $t$, and $\bm\Omega_{\mathrm Q}^{\mathrm{time}}$, $\bm\Omega_{\mathrm K}^{\mathrm{time}}$, and $\bm\Omega_{\mathrm V}^{\mathrm{time}}$ are the trainable query, key, and value projection matrices of T-ATT, respectively. The output $\bm u_{k,t}\in\mathbb{R}^d$ summarizes the multimodal history of user $k$ over the $H$ frames. Since the poses are designed for the next $F$ time steps, we use $\bm u_{k,t}$ to predict the future features of user $k$ as
\begin{equation}
\bm u_{k,t,f}=f_{\mathrm{future}}(\bm u_{k,t},f\Delta_{\mathrm s},F\Delta_{\mathrm s}),\quad f\in\mathcal F,
\label{eq:future}
\end{equation}
where $f_{\mathrm{future}}$ is an MLP shared by all users, and $\bm u_{k,t,f}\in\mathbb{R}^d$ is the feature of user $k$ predicted at time step $t$ for time step $t+f$. The inputs $f\Delta_{\mathrm s}$ and $F\Delta_{\mathrm s}$ are the durations from time step $t$ to time step $t+f$ and to the last designed time step $t+F$, respectively, as shown in Fig.~\ref{fig:timeline}.
\subsection{Receding-Horizon 6DMA Trajectory Design}\label{sec:trajectory}
The kinematic constraints \eqref{eq:step_position} and \eqref{eq:step_rotation} couple consecutive poses while the users and objects keep moving, so that selecting the poses for the next time step alone may leave favorable later poses out of reach. To account for this coupling, we design the 6DMA trajectory $\bm x_{t,1:F}$ for the next $F$ time steps, and only its first step is executed as in \eqref{eq:execute}.

Since a surface provides a high gain only for the users whose paths it faces, S-ATT weights the predicted features of the users for each surface according to its previous pose. As shown by the loop in Fig.~\ref{fig:architecture}, we generate the trajectory sequentially, starting from $\bm x_{t,0}=\bm x_t$. At step $f$, an MLP $f_{\mathrm Q}$ forms the query input of surface $b$ from its center position $\bm{q}_{b,t,f-1}$ and rotation $\bm{R}_{b,t,f-1}$ at the previous step, and the predicted features of the users form the keys and values. The output of S-ATT for surface $b$ at step $f$ is given by
\begin{equation}
\bm\varrho_{b,t,f}=\sum_{k\in\mathcal K}\zeta_{b,k,t,f}^{\mathrm{sur}}\bm\Omega_{\mathrm V}^{\mathrm{sur}}\bm u_{k,t,f},
\label{eq:satt}
\end{equation}
with the attention weight being
\begin{align}
\zeta_{b,k,t,f}^{\mathrm{sur}}&=\frac{\exp(\epsilon_{b,k,t,f}^{\mathrm{sur}})}{\sum_{k'\in\mathcal K}\exp(\epsilon_{b,k',t,f}^{\mathrm{sur}})},\label{eq:satt_weight_w}\\
\epsilon_{b,k,t,f}^{\mathrm{sur}}&=\frac{\big(\bm\Omega_{\mathrm Q}^{\mathrm{sur}}f_{\mathrm Q}(\bm{q}_{b,t,f-1},\bm{R}_{b,t,f-1})\big)^{\mathsf T}\bm\Omega_{\mathrm K}^{\mathrm{sur}}\bm u_{k,t,f}}{\sqrt{d_{\mathrm h}}},\label{eq:satt_weight}
\end{align}
where $\epsilon_{b,k,t,f}^{\mathrm{sur}}$ is the attention score, $\zeta_{b,k,t,f}^{\mathrm{sur}}$ measures the importance of user $k$ to surface $b$ at step $f$, and $\bm\Omega_{\mathrm Q}^{\mathrm{sur}}$, $\bm\Omega_{\mathrm K}^{\mathrm{sur}}$, and $\bm\Omega_{\mathrm V}^{\mathrm{sur}}$ are the trainable query, key, and value projection matrices of S-ATT, respectively. An MLP decoder, shared by all surfaces, then turns this user information into a candidate pose of surface $b$. The decoder maps $\bm\varrho_{b,t,f}$, the poses of all surfaces at step $f-1$, and the speed limits $v_{\max}$ and $\omega_{\max}$ to the candidate position $\widetilde{\bm{q}}_{b,t,f}$ and the candidate Euler angles of surface $b$ at step $f$, which give the candidate rotation $\widetilde{\bm{R}}_{b,t,f}$ by \eqref{eq:rotation_matrix}. Since the candidate pose may not satisfy the geometric constraints or the kinematic constraints, we apply the projection layer, and the projected pose of step $f$ forms the query of step $f+1$.

\subsection{Differentiable Projection Layer}\label{sec:recovery}
The decoder can output any pose, whereas the surfaces have to satisfy the geometric constraints and move and rotate by at most the distance and the angle that their motors allow within one time step. To satisfy constraints \eqref{eq:box}--\eqref{eq:step_rotation}, the projection layer maps the candidate pose of surface $b$ at step $f$ to a feasible pose, starting from the feasible pose of the previous step. We first project the candidate position onto the site space $\mathcal C$, i.e., we replace it with its closest point $\bm{q}_{b,t,f}^{\mathcal C}$ in $\mathcal C$. For a cubic site space, this projection clips each coordinate of the candidate position to the cube. Then, we scale the translation and the rotation to meet the speed limits as
\begin{align}
\bm{q}_{b,t,f}&=\bm{q}_{b,t,f-1}+\eta_{b,t,f}(\bm{q}_{b,t,f}^{\mathcal C}-\bm{q}_{b,t,f-1}),\label{eq:bridge}\\
\bm{R}_{b,t,f}&=\mathrm{Exp}_{\mathrm{SO}(3)}\!\big(\eta_{b,t,f}^{\mathrm R}\,\mathrm{Log}_{\mathrm{SO}(3)}(\widetilde{\bm{R}}_{b,t,f}\bm{R}_{b,t,f-1}^{\mathsf T})\big)\bm{R}_{b,t,f-1},\label{eq:bridge_rotation}
\end{align}
where $\eta_{b,t,f}=\min\{1,v_{\max}\Delta_{\mathrm s}/\lVert \bm{q}_{b,t,f}^{\mathcal C}-\bm{q}_{b,t,f-1}\rVert_2\}$ and $\eta_{b,t,f}^{\mathrm R}=\min\{1,\omega_{\max}\Delta_{\mathrm s}/d_{\mathrm R}(\bm{R}_{b,t,f-1},\widetilde{\bm{R}}_{b,t,f})\}$ are the scaling factors of the translation and the rotation of surface $b$ at step $f$, respectively. The update in \eqref{eq:bridge} moves the surface center along a straight line by at most $v_{\max}\Delta_{\mathrm s}$, and the new position stays in $\mathcal C$ since $\mathcal C$ is convex. In \eqref{eq:bridge_rotation}, the logarithm map $\mathrm{Log}_{\mathrm{SO}(3)}(\cdot)$ and the exponential map $\mathrm{Exp}_{\mathrm{SO}(3)}(\cdot)$ convert the relative rotation into its axis and angle and back, so that the surface rotates toward $\widetilde{\bm{R}}_{b,t,f}$ about this axis \mbox{by at most $\omega_{\max}\Delta_{\mathrm s}$.}

The new poses may still not satisfy constraints \eqref{eq:separation}--\eqref{eq:outward}. In this case, let $\mathcal B_{t,f}^{\mathrm v}\subseteq\mathcal B$ denote the set of surfaces involved in a violated constraint at step $f$. We halve the scaling factors $\eta_{b,t,f}$ and $\eta_{b,t,f}^{\mathrm R}$ of each surface $b\in\mathcal B_{t,f}^{\mathrm v}$, i.e., the surface moves and rotates only half as far toward its candidate pose, recompute its pose by \eqref{eq:bridge} and \eqref{eq:bridge_rotation}, check constraints \eqref{eq:separation}--\eqref{eq:outward} again, and update $\mathcal B_{t,f}^{\mathrm v}$. If a surface is still involved in a violated constraint after its scaling factors have been halved a predefined number of times, we set them to zero, so that it keeps its pose of the previous step. Since each surface involved in a newly violated constraint joins $\mathcal B_{t,f}^{\mathrm v}$, in the worst case, all surfaces keep their previous poses. Hence, the projected poses are feasible if the poses of the previous step are feasible. Since each designed trajectory starts from the feasible current poses $\bm x_t$, its poses satisfy the geometric constraints \eqref{eq:box}--\eqref{eq:outward} at every time step and the kinematic constraints \eqref{eq:step_position} and \eqref{eq:step_rotation} of problem~\eqref{eq:problem}. Moreover, the projected pose obtained by \eqref{eq:bridge} and \eqref{eq:bridge_rotation} is differentiable with respect to the candidate pose almost everywhere. During end-to-end training, the projection layer thus allows the gradients to be backpropagated from the objective to the decoder.

\subsection{Pilot-Based Beamforming}\label{sec:beamforming}
After the surfaces reach the poses $\bm x_t$, the BS designs the beamformers based on these poses, the pilots received at these poses, and the features of time step $t-1$. Each beamformer $\bm{w}_{k,t}=\sqrt{p_{k,t}}\bm{d}_{k,t}$ consists of a unit-norm beamforming direction $\bm{d}_{k,t}$ and the power $p_{k,t}$ allocated to user $k$ along this direction. We obtain the directions from the channel estimates by RZF, and we learn the power allocation by P-ATT, as described below.

For the directions, let $\widehat{\bm H}_t\triangleq[\widehat{\bm{h}}_{1,t},\ldots,\widehat{\bm{h}}_{K,t}]\in\mathbb{C}^{N\times K}$ denote the matrix of the channel estimates at the reached poses. The RZF beamforming matrix is $\sqrt{P_{\mathrm T}}\bm V_t/\lVert \bm V_t\rVert_{\mathrm F}$, where $\bm V_t$ is the unnormalized matrix given by
\begin{equation}
\bm V_t\triangleq\widehat{\bm H}_t\left(\widehat{\bm H}_t^{\mathsf H}\widehat{\bm H}_t
+\frac{K\sigma^2}{P_{\mathrm T}}\bm I_K\right)^{-1},
\label{eq:earzf}
\end{equation}
and the beamforming direction for user $k$ is the normalized column $\bm{d}_{k,t}\triangleq[\bm V_t]_{:,k}/\lVert [\bm V_t]_{:,k}\rVert_2$.

For the power allocation, let $\widehat G_{k,i,t}\triangleq|\widehat{\bm{h}}_{k,t}^{\mathsf H}\bm{d}_{i,t}|^2$ denote the estimated gain of the beamforming direction $\bm{d}_{i,t}$ of user $i$ at user $k$. Here, $\widehat G_{k,k,t}$ is the desired signal gain of user $k$, and $\widehat G_{k,i,t}$ with $i\neq k$ is the undesired interference gain from user $i$ to user $k$. With the beamforming directions fixed, problem~\eqref{eq:bf_sub} under the estimated gains reduces to a power allocation problem. By introducing an auxiliary variable $\Gamma_{\min}$ for the minimum SINR, we can express this problem as
\begin{subequations}\label{eq:maxmin_power}
\begin{align}
\underset{\Gamma_{\min},\,p_{1,t},\ldots,p_{K,t}}{\text{maximize}}\ &\Gamma_{\min}\\
\text{subject to}\ &\frac{p_{k,t}\widehat G_{k,k,t}}{\sigma^2+\sum_{i\in\mathcal K\setminus\{k\}}p_{i,t}\widehat G_{k,i,t}}\geq\Gamma_{\min},\ k\in\mathcal K,\\
&\sum_{k\in\mathcal K}p_{k,t}\leq P_{\mathrm T},\label{eq:maxmin_power_budget}\\
&p_{k,t}\geq0,\ k\in\mathcal K.
\end{align}
\end{subequations}
With a short pilot length, the solution to problem~\eqref{eq:maxmin_power} suffers from the estimation error of $\widehat G_{k,i,t}$. As discussed in Section~\ref{sec:motivation}, the multimodal sensing data help reduce this degradation. Hence, we feed the estimated gains, the multimodal features, and the current poses into P-ATT, and train it to maximize the minimum rate under the true channels.

To capture the multiuser interference, inspired by the graph-based designs in \cite{Eisen2020REGNN,11062505}, we model the users as the $K$ nodes of a graph whose edges carry the interference gains, with self-loops that carry the desired signal gains. This interference graph is directed, since the interference gain $\widehat G_{k,i,t}$ from user $i$ to user $k$ generally differs from the interference gain $\widehat G_{i,k,t}$ from user $k$ to user $i$. On this graph, P-ATT combines, for each user, its desired signal gain, the interference gains from the other users, its multimodal feature, and the current poses to determine its power. P-ATT is a graph neural network (GNN) with $N_{\mathrm g}$ layers in the message-passing form, i.e., in each layer, each node aggregates the information from all nodes with attention weights, as in a graph attention network, and combines it with its own feature. Let $\bm\nu_{k,t}^{(l)}\in\mathbb{R}^d$ denote the feature of node $k$ at layer $l$, where the initial feature $\bm\nu_{k,t}^{(0)}=f_{\mathrm{in}}(P_{\mathrm T}\widehat G_{k,k,t}/\sigma^2,\bm u_{k,t-1},\bm x_t)$ is obtained by an MLP $f_{\mathrm{in}}$. The normalization of $\widehat G_{k,k,t}$ follows from the SINR in \eqref{eq:sinr}. With $\bm{w}_{k,t}=\sqrt{p_{k,t}}\bm{d}_{k,t}$, the SINR depends only on the normalized gains $P_{\mathrm T}|\bm{h}_{k,t}^{\mathsf H}(\bm x_t)\bm{d}_{i,t}|^2/\sigma^2$ and the power fractions $p_{i,t}/P_{\mathrm T}$, $i\in\mathcal K$. P-ATT outputs these power fractions. Layer $l$ updates the feature of node $k$ as
\begin{equation}
\bm\nu_{k,t}^{(l+1)}=f_{\mathrm n}\!\left(\bm\nu_{k,t}^{(l)},
\sum_{i\in\mathcal K}\zeta_{k,i,t}^{(l)}f_{\mathrm v}(\bm\nu_{i,t}^{(l)},\bm\kappa_{k,i,t})\right),
\label{eq:graph_attention}
\end{equation}
where $\bm\kappa_{k,i,t}$ is the feature of the edge from node $i$ to node $k$ (a self-loop if $i=k$), obtained from the estimated gain $\widehat G_{k,i,t}$ normalized as $P_{\mathrm T}\widehat G_{k,i,t}/\sigma^2$. The MLP $f_{\mathrm v}$, shared by all pairs of nodes, maps the feature $\bm\nu_{i,t}^{(l)}$ of node $i$ and the edge feature $\bm\kappa_{k,i,t}$ to the message from node $i$ to node $k$. The MLP $f_{\mathrm n}$, shared by all nodes, combines the weighted sum of the messages with the previous feature $\bm\nu_{k,t}^{(l)}$ of node $k$. The attention weight $\zeta_{k,i,t}^{(l)}$, which measures the importance of node $i$ to node $k$ at layer $l$, is given by
\begin{align}
\zeta_{k,i,t}^{(l)}&=\frac{\exp(\epsilon_{k,i,t}^{(l)})}{\sum_{i'\in\mathcal K}\exp(\epsilon_{k,i',t}^{(l)})},\label{eq:patt_weight_w}\\
\epsilon_{k,i,t}^{(l)}&=\frac{(\bm\Omega_{\mathrm Q}^{(l)}\bm\nu_{k,t}^{(l)})^{\mathsf T}\bm\Omega_{\mathrm K}^{(l)}\bm\nu_{i,t}^{(l)}}{\sqrt{d_{\mathrm h}}}+f_{\mathrm e}(\bm\kappa_{k,i,t}),\label{eq:patt_weight}
\end{align}
where $\epsilon_{k,i,t}^{(l)}$ is the attention score, $\bm\Omega_{\mathrm Q}^{(l)}$ and $\bm\Omega_{\mathrm K}^{(l)}$ are the trainable query and key projection matrices of layer $l$, respectively, and $f_{\mathrm e}$ is an MLP that maps the edge feature to a scalar. We include $f_{\mathrm e}(\bm\kappa_{k,i,t})$ in the attention score $\epsilon_{k,i,t}^{(l)}$, so that the weight that node $k$ gives to node $i$ also depends on the estimated gain $\widehat G_{k,i,t}$ \cite{ying2021graphormer}. After the last layer, we map the feature $\bm\nu_{k,t}^{(N_{\mathrm g})}$ of node $k$ to a score $\chi_{k,t}$ by an MLP shared by all nodes. To satisfy the transmit power constraint \eqref{eq:power}, we normalize the scores jointly across all users by a softmax activation function. The power allocated to user $k$ is $p_{k,t}=P_{\mathrm T}\exp(\chi_{k,t})/\sum_{i\in\mathcal K}\exp(\chi_{i,t})$.
\subsection{Training Objective and Online Inference}\label{sec:training}
In the training, we evaluate each designed trajectory and its beamformers by the minimum rate on the simulated future channels at the designed poses. We train the network on $N_{\mathrm{tr}}$ samples. Each sample $r$ is taken at a time step $t_r$ and contains the feasible current 6D poses $\bm x_{t_r}$, the multimodal sensing data $\mathcal I_{t_r+f-1}$ in \eqref{eq:information} for $f\in\mathcal F$, and the propagation environment at time steps $t_r+1,\ldots,t_r+F$. For sample $r$, the trajectory design module outputs the 6DMA trajectory $\bm x_{t_r,1:F}=\pi_{\bm\theta}(\mathcal I_{t_r},\bm x_{t_r})$. We simulate the channels at the poses $\bm x_{t_r,f}$ at time steps $t_r+f$ in the scenes of Section~\ref{sec:results} using Sionna RT. At each designed step $f\in\mathcal F$, we generate the pilots $\bm Y_{t_r,f}$ at the poses $\bm x_{t_r,f}$ by \eqref{eq:pilot}, and the beamforming mapping outputs the beamformers $\bm{W}_{t_r,f}=\mathcal W_{\bm\theta,\bm\psi}(\mathcal I_{t_r+f-1},\bm x_{t_r,f},\bm Y_{t_r,f})$, which achieve the rate $R_{k,t_r+f}(\bm x_{t_r,f},\bm{W}_{t_r,f})$ for user $k$ at time step $t_r+f$. Since the channels computed by Sionna RT are differentiable with respect to the positions and rotations of the surfaces, the gradients are backpropagated through the computed channels to the network outputs. We minimize the loss function
\begin{equation}
\begin{split}
\mathcal J(\bm\theta,\bm\psi)={}&-\frac{1}{N_{\mathrm{tr}}F}\sum_{r=1}^{N_{\mathrm{tr}}}\sum_{f\in\mathcal F}\min_{k\in\mathcal K}R_{k,t_r+f}(\bm x_{t_r,f},\bm{W}_{t_r,f})\\
&+\mu\,\mathcal J_{\mathrm{pen}},
\end{split}
\label{eq:loss}
\end{equation}
where the first term is the negative minimum rate averaged over the designed future time steps on the training set and serves as a surrogate for the negative of the objective in problem~\eqref{eq:problem}, and $\mu>0$ is the penalty coefficient. Although only the first designed step is executed online, the loss over all $F$ designed steps favors a first step that keeps favorable later poses reachable. The second term in \eqref{eq:loss} is the penalty term, which accounts for constraints \eqref{eq:box}--\eqref{eq:step_rotation} in the training. It measures the distance between each candidate pose and the corresponding projected pose, and is given by
\begin{equation}
\begin{split}
\mathcal J_{\mathrm{pen}}=\frac{1}{N_{\mathrm{tr}}FB}\sum_{r=1}^{N_{\mathrm{tr}}}\sum_{f\in\mathcal F}\sum_{b\in\mathcal B}\bigg(&\frac{\lVert \widetilde{\bm{q}}_{b,t_r,f}-\bm{q}_{b,t_r,f}\rVert_2^2}{(v_{\max}\Delta_{\mathrm s})^2}\\
&+\frac{d_{\mathrm R}^2(\widetilde{\bm{R}}_{b,t_r,f},\bm{R}_{b,t_r,f})}{(\omega_{\max}\Delta_{\mathrm s})^2}\bigg),
\end{split}
\label{eq:rec_loss}
\end{equation}
where $\widetilde{\bm{q}}_{b,t_r,f}$ and $\widetilde{\bm{R}}_{b,t_r,f}$ are the candidate position and rotation of surface $b$ at step $f$ of sample $r$, respectively, $\bm{q}_{b,t_r,f}$ and $\bm{R}_{b,t_r,f}$ are their projections, and the denominators are the squared maximum position and rotation changes per time step, respectively. The penalty term pulls each candidate pose that does not satisfy the constraints toward its projected pose. During online inference, the BS applies the trained $\pi_{\bm\theta}$ and $\mathcal W_{\bm\theta,\bm\psi}$ at each time step as in \eqref{eq:plan}--\eqref{eq:beammap}.
\section{Performance Evaluation}
\label{sec:results}
In this section, we evaluate the proposed design by simulations and compare it with the state of the art. Since the existing multimodal sensing-aided beamforming designs consider FPAs, we compare them with the proposed beamforming on the same FPA, which shows the gain of the proposed beamforming over these designs. We also compare with 6DMA schemes that use fewer sensing modalities or replace one attention module, which shows the contribution of each part of the proposed design.

\subsection{Simulation Setup}
\begin{figure*}[!t]
\centering\includegraphics[width=1.0\textwidth]{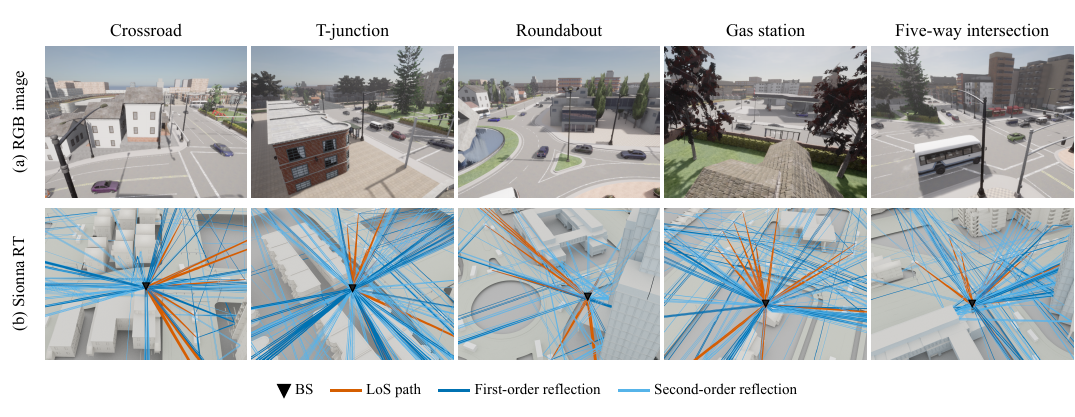}
\caption{The five scenes of the digital twin, where (a) shows the RGB images from CARLA and (b) shows the propagation paths from Sionna RT.}
\label{fig:scenes}
\end{figure*}
To train and evaluate the proposed design, we follow the approach in \cite{Charan2021Blockage,LiYu2026CameraLiDARRF}, which generates the sensing data and the ray-traced channels from the same simulated scene. We build a digital twin of five traffic scenes with the CARLA simulator \cite{2017arXiv171103938D} and Sionna RT \cite{10465179}. At each time step, CARLA simulates the traffic and generates the RGB images, the LiDAR point clouds, and the GPS positions. Sionna RT computes the propagation paths on the 3D models of the same scene, including the vehicles at their current positions. The five scenes are in the Town03 map of CARLA, namely a crossroad, a \mbox{T-junction}, a roundabout, a gas station, and a five-way intersection, as shown in Fig.~\ref{fig:scenes}.

Unless stated otherwise, we consider the following default setup. The BS is equipped with $B=4$ surfaces, each being a $4\times4$ uniform planar array with half-wavelength spacing, i.e., $N_{\mathrm{sur}}=16$ and $N=64$, and serves $K=16$ users. The site space is a cube with a side length of $1$~m, and the minimum distance between the surface centers is $d_{\min}=0.25$~m. In the antenna radiation pattern in \eqref{eq:pattern}, we set $\vartheta_{\mathrm{3dB}}=\varphi_{\mathrm{3dB}}=65^\circ$, $G_{\max}=8$~dBi, and $A_{\mathrm m}=30$~dB. The carrier frequency is $3.5$~GHz, and the noise power at the BS and the users is $-87$~dBm. We set the transmit power budget to $P_{\mathrm T}=46$~dBm \cite{massiveBook}, the pilot power to $P_{\mathrm p}=23$~dBm \cite{Ji2026SingleStep6DMA}, the pilot length to $\tau_{\mathrm p}=16$, and the history length and the horizon to $H=F=5$. The translation and rotation speed limits are $v_{\max}=0.5$~m/s and $\omega_{\max}=0.5$~rad/s, respectively, and the interval between consecutive time steps is $\Delta_{\mathrm s}=100$~ms.

For the sensing, the BS is equipped with three RGB cameras and a LiDAR sensor. Each RGB camera has a resolution of $960\times540$ pixels and a horizontal field of view (FoV) of $110^\circ$, and adjacent cameras are $40^\circ$ apart. The LiDAR sensor has $64$ laser beams, a horizontal FoV of $360^\circ$, and a range of $120$~m.

For the neural network, we set the feature dimension to $d=256$, the number of graph-attention layers in P-ATT to $N_{\mathrm g}=4$, and the penalty coefficient in \eqref{eq:loss} to $\mu=0.1$. We train the network using the AdamW optimizer with a learning rate of $10^{-4}$ and a batch size of $32$ for $100$ epochs. We report the minimum rate in bit/s/Hz at the reached 6D poses, averaged over the test time steps.

\subsection{Effect of Multimodal Sensing}
We first compare the proposed design with the following baselines. All schemes use the same uplink pilots.
\begin{itemize}
\item \textbf{6DMA with RGB+GPS} and \textbf{6DMA with LiDAR+GPS}: These baselines remove the LiDAR input and the RGB input, respectively, from the multimodal history encoding, whose features feed the trajectory design and P-ATT. Their gaps to the proposed design measure the gains of the LiDAR and RGB inputs, respectively.
\item \textbf{6DMA with GPS only}: This baseline uses only the GPS positions as the sensing input of the multimodal history encoding and thus observes only the users. Its gap to the proposed design measures the gain of the RGB and LiDAR inputs.
\item \textbf{FPA with proposed beamforming}: This baseline keeps the surfaces at the fixed poses of the four-sector FPA in \cite{Ji2026CSIFree6DMAIoV} and uses the proposed beamforming. Its gap to the proposed design measures the gain of 6DMA.
\item \textbf{FPA with RGB+LiDAR GNN}: This baseline uses the FPA defined above and the network in \cite{LiYu2026CameraLiDARRF} to generate the beamformers. The inputs of this network are the LS channel estimates in \eqref{eq:ls} and the LiDAR points of the users and objects detected in the RGB images. Its gap to FPA with proposed beamforming measures the gain of the proposed beamforming over this multimodal beamforming design on the same FPA.
\item \textbf{FPA with two-stage beamforming}: This baseline uses the FPA defined above with the two-stage scheme in \cite{PatelHeath2024}. In the first stage, it aligns one beam with each user based on the RGB images, the LiDAR point clouds, and the GPS positions. In the second stage, it obtains the digital beamformers by RZF from the estimated $K\times K$ effective channel of the aligned beams. Its gap to FPA with proposed beamforming measures the gain of the proposed beamforming over this two-stage design.
\end{itemize}

\begin{figure}[!t]
\centering\includegraphics[width=\columnwidth]{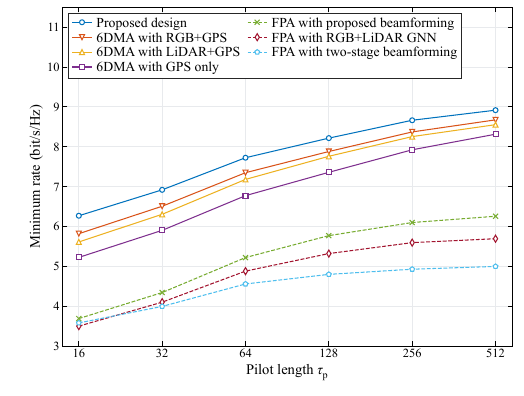}
\caption{Minimum rate versus the pilot length.}
\label{fig:pilot}
\end{figure}
Fig.~\ref{fig:pilot} shows the minimum rate versus the pilot length. We observe that the proposed design achieves a higher minimum rate than the other 6DMA schemes and FPA with proposed beamforming for all pilot lengths. When compared with 6DMA with GPS only, 6DMA with RGB+GPS and 6DMA with LiDAR+GPS improve the minimum rate by $11\%$ and $7\%$ at $\tau_{\mathrm p}=16$, respectively, and the proposed design with all three modalities improves it by $20\%$. This is because the RGB images and the LiDAR point clouds capture the objects in the propagation environment, whereas the GPS positions do not reveal these objects. The RGB images contribute more than the LiDAR point clouds, since the RGB images show the classes of the objects, such as large vehicles. Moreover, adding the LiDAR point clouds to 6DMA with RGB+GPS still raises the minimum rate by $8\%$, since the point clouds capture the 3D geometry that the RGB images lack. As the pilot length grows, the gain of the proposed design over 6DMA with GPS only decreases from $20\%$ at $\tau_{\mathrm p}=16$ to $7\%$ at $\tau_{\mathrm p}=512$. This is because the RGB images and the LiDAR point clouds help P-ATT compensate for the channel estimation error, which decreases with the pilot length. The remaining gain at $\tau_{\mathrm p}=512$ thus stems mainly from the poses designed with the RGB and LiDAR inputs.

When compared with FPA with proposed beamforming, the proposed design improves the minimum rate by $70\%$ and $43\%$ at $\tau_{\mathrm p}=16$ and $512$, respectively. This is because the 6DMA surfaces move and rotate to face the strongest paths of the users, which raises the effective antenna gains in \eqref{eq:pattern} and, at a short pilot length, also reduces the relative estimation error in \eqref{eq:ls}. Among the FPA schemes, FPA with RGB+LiDAR GNN is $5\%$ lower than FPA with proposed beamforming at $\tau_{\mathrm p}=16$, and the gap increases to $9\%$ at $\tau_{\mathrm p}=512$. This is because its network learns the beamformers of all $K$ users over all $N$ antennas, whereas the proposed beamforming obtains the directions from the channel estimates by RZF and learns only the $K$ powers. As the estimation error decreases, the RZF directions suppress the multiuser interference more effectively. The gap of FPA with two-stage beamforming to FPA with proposed beamforming increases from $3\%$ at $\tau_{\mathrm p}=16$ to $20\%$ at $\tau_{\mathrm p}=512$. This is because its digital beamformers act on the $K$ aligned beams instead of all $N$ antennas, and the errors of the sensing-based beam alignment do not decrease with the pilot length.

\subsection{Effect of the Proposed Modules}
\begin{figure}[!t]
\centering\includegraphics[width=\columnwidth]{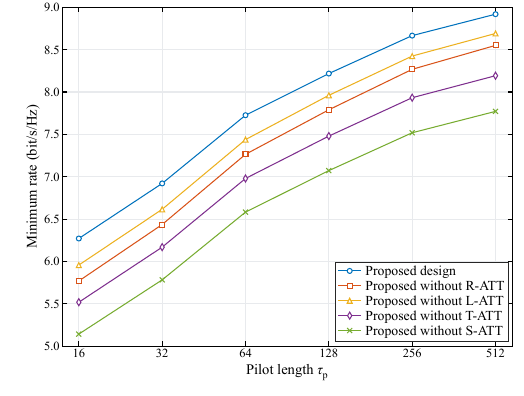}
\caption{Minimum rate versus the pilot length in the ablation study.}
\label{fig:ablation}
\end{figure}
\begin{figure}[!t]
\centering\includegraphics[width=\columnwidth]{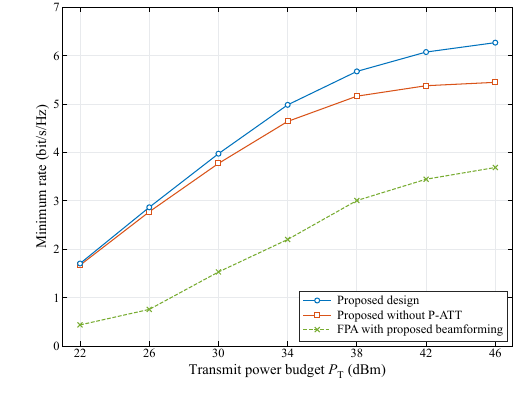}
\caption{Minimum rate versus the transmit power budget.}
\label{fig:power}
\end{figure}
\begin{figure*}[!t]
\centering\includegraphics[width=\textwidth]{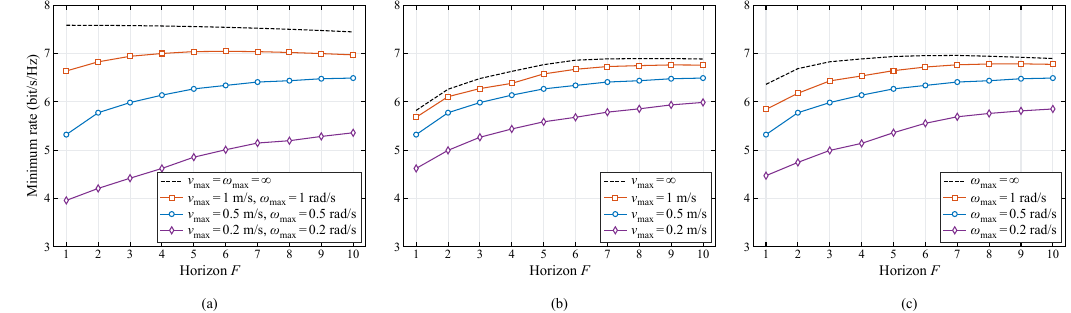}
\caption{Minimum rate versus the horizon $F$, where (a) both $v_{\max}$ and $\omega_{\max}$ vary, (b) only $v_{\max}$ varies, and (c) only $\omega_{\max}$ varies.}
\label{fig:horizon}
\end{figure*}
In Figs.~\ref{fig:ablation} and \ref{fig:power}, we compare our proposed design with the following baselines.
\begin{itemize}
\item \textbf{Proposed without R-ATT} and \textbf{Proposed without L-ATT}: These baselines replace R-ATT and L-ATT, respectively, with an MLP that takes the pooled features of the corresponding modality concatenated with the GPS feature. Their gaps to the proposed design measure the respective gains of the user-specific RGB \mbox{and LiDAR features.}
\item \textbf{Proposed without T-ATT}: This baseline replaces T-ATT with an MLP over the concatenated fused features of the $H$ frames. Its gap to the proposed design measures the gain of weighting the sensing frames by attention.
\item \textbf{Proposed without S-ATT}: This baseline replaces S-ATT with an MLP that maps the concatenated features of all users and the previous pose of each surface to $\bm\varrho_{b,t,f}$. Its gap to the proposed design measures the gain of weighting the users by attention for each surface.
\item \textbf{Proposed without P-ATT}: This baseline adopts the RZF beamforming matrix based on $\bm V_t$ in \eqref{eq:earzf} instead of P-ATT. Its gap to the proposed design measures the gain of P-ATT.
\end{itemize}

Fig.~\ref{fig:ablation} shows the minimum rate versus the pilot length in the ablation study. We observe that replacing R-ATT, L-ATT, T-ATT, or S-ATT reduces the minimum rate for all pilot lengths. At $\tau_{\mathrm p}=16$, replacing S-ATT, T-ATT, R-ATT, and L-ATT reduces the minimum rate by $18\%$, $12\%$, $8\%$, and $5\%$, respectively. S-ATT is more important than the other attention modules, since it selects, regardless of the user order, the users whose paths each surface can face, whereas its replacement has to learn this selection from features concatenated in an arbitrary user order. T-ATT weights the sensing frames by their relevance for predicting the user movement, whereas its replacement lacks explicit attention weights for selecting the sensing frames. R-ATT and L-ATT weight the scene features for each user according to the objects in its propagation environment, whereas their replacements give every user the same pooled scene features. As the pilot length grows, the minimum-rate losses of the proposed design without R-ATT and the proposed design without L-ATT decrease from $0.50$ and $0.31$~bit/s/Hz at $\tau_{\mathrm p}=16$ to $0.37$ and $0.23$~bit/s/Hz at $\tau_{\mathrm p}=512$, respectively. This is because R-ATT and L-ATT bring the scene information around each user into the features of P-ATT, and this information matters less when the channel estimates are accurate. Replacing R-ATT leads to a larger loss than replacing L-ATT, which is consistent with the larger contribution of the RGB images in Fig.~\ref{fig:pilot}.

Fig.~\ref{fig:power} shows the minimum rate versus the transmit power budget. We observe that P-ATT improves the minimum rate over the proposed design without P-ATT for all transmit power budgets, and that the gain increases from $2\%$ at $22$~dBm to $15\%$ at $46$~dBm. This is because the rates are noise-limited at a low transmit power budget and interference-limited at a high transmit power budget, where the interference caused by the estimation error grows as fast as the signal power. P-ATT allocates the power based on the estimated gains, the multimodal features, and the poses to raise the minimum rate, whereas the proposed design without P-ATT fixes the power allocation by $\bm V_t$ in \eqref{eq:earzf}. The proposed design also outperforms FPA with proposed beamforming for all transmit power budgets.
\subsection{Effect of Receding-Horizon Design}
Finally, Fig.~\ref{fig:horizon} shows the minimum rate versus the horizon $F$. As idealized references, the dashed lines show the proposed design trained and evaluated without the corresponding kinematic constraints. For the dashed line in Fig.~\ref{fig:horizon}(a), the projection layer enforces only the geometric constraints.

In Fig.~\ref{fig:horizon}, we observe that the gain from increasing $F$ grows as the kinematic constraints become more stringent. Under the default constraints, the minimum rate increases by $18\%$ from $F=1$ to $F=5$ and by $22\%$ from $F=1$ to $F=10$ in Fig.~\ref{fig:horizon}(a). This is because moving a surface to a favorable pose takes several time steps, and a longer horizon lets the surface move earlier toward a pose that becomes favorable later. When $v_{\max}$ and $\omega_{\max}$ are reduced to $0.2$~m/s and $0.2$~rad/s, respectively, the gain from $F=1$ to $F=10$ increases to $35\%$, since a pose change then takes more time steps. When they are relaxed to $1$~m/s and $1$~rad/s, respectively, the minimum rate increases by $6\%$ up to $F=6$ and then decreases slightly. This is because under loose constraints, a pose change takes fewer time steps, so that the later designed steps bring little gain, but the less accurate predictions for these steps degrade the executed first step. In Fig.~\ref{fig:horizon}(a), the dashed line decreases slightly with $F$, since with infinite speed, the surfaces can reach any feasible pose within one time step and the later designed steps only add prediction errors. In Figs.~\ref{fig:horizon}(b) and \ref{fig:horizon}(c), the dashed lines still increase with $F$ up to about $F=7$, since the remaining constraint slows down pose changes.
\section{Conclusion}
In this paper, we proposed a multimodal sensing-aided trajectory and beamforming design for multiuser downlink 6DMA systems. The BS used the RGB, LiDAR, and GPS histories to design the 6DMA trajectory over several time steps under the kinematic constraints, executed only the first step, and designed the beamformers from the pilots at the reached poses. Simulation results showed minimum-rate gains of the proposed design over FPAs, and ablation studies quantified the contributions of multimodal sensing and the proposed attention modules. The proposed beamforming also achieved a higher minimum rate than two existing multimodal beamforming designs. Designing several time steps ahead was more beneficial under tighter speed limits of the surfaces. These results show that observing the users and the surrounding objects and designing several time steps ahead help a 6DMA-BS select favorable poses before the channels at these poses can be estimated.

\end{document}